# Information and evaluation models of complex hierarchical network systems

Olexandr Polishchuk, Mykhailo Yadzhak

*Pidstryhach Institute for Applied Problems of Mechanics and Mathematics, National Academy of Sciences of Ukraine, Naukova str., 3b, Lviv, 79060, Ukraine*

**Abstract**

The problem of selection, storage, search and analysis of information about the state, functioning and interaction of elements of complex hierarchical network systems is considered. The principles of construction of information models of such systems and models of evaluation of their operation efficiency are proposed. These models are formed on the basis of results of regular planned studies of the system and continuous monitoring of the activities of its components at all levels of the hierarchy. The problem of novelty detection in the information model of system is investigated and it is shown how the application of evaluation theory methods allows to solve it. The proposed methods of complex analysis of the behavior of real complex systems and optimization of data processing makes it possible to significantly simplify and speed up the decision-making process for further action for such systems.

## 1. Introduction

Our understanding about the world around us is formed on the basis of available information about it. Data can be obtained with the help of visual and special means of observation, experimental and theoretical research, etc. According to the International Data Corporation, the amount of data created by humanity in 2020 was approximately $4\times10^{10}$ GB [1]. The amount of useful information, ie data, the analysis of which can improve the state and operation process of man-made systems of different types and purposes according to a predetermined set of criteria and parameters, according to various estimates is in the range of 23-35%, and the volume of data, which can be analyzed and used to optimize the operation of such systems – within 1-3% of their total amount [2]. Thus, of all the experimental data obtained at the Large Andron Collider (Cern), no more than 0.004% is analyzed (the cost of building the collider was more than 4.6 billion euros) [3]. It is obvious that human is interested in improving the efficiency of each system created by him. Theoretically, this problem is quite simple to solve: authorized person need to receive in a timely manner all necessary useful information about the system and on the basis of its analysis to make appropriate management decisions. In practice, it is rare that a manager who makes these decisions is completely satisfied with the timeliness, synchronization and quality of data received by him [4, 5]. This is usually due to too large amounts of information circulating in the system and placed in many information systems and data warehouses [6, 7]. If earlier during the development of decision support means the main attention was paid to solving the problem of uncertainty caused by the lack of complete data about the system [8, 9], then with the advent of the third millennium and the rapid development of new information technologies there was a problem of decision-making in the conditions of information overload caused by necessity of processing of too big volumes of data [10, 11]. But the main problem is not so much in the amount of data, but in the non-availability of effective methods for identifying, processing and analyzing useful information [12]. At present, this problem is usually solved by developing methods for finding the necessary information in poorly structured and "littered" with various "outliers" (ie erroneous, unimportant, inaccurate and duplicate information) data warehouses. This is the simplest, but not the most effective way to solve this problem, especially given that over time, the rate of increase of information will continue to grow. An alternative approach is to form structured information models (IM) and evaluation models (EM) for at least the most important for human society real complex systems (medical, educational, environmental, security [13, 14], etc.), in which the identification and elimination more " outliers" occur at the filling stage, and therefore the search for useful information is much faster, easier and more efficient. Equally important is the minimization of subjective influences during the selection and processing of empirical data about the system to obtain objective and mathematically sound conclusions about its state, operation efficiency and interaction with the environment.

The purpose of article is to determine the principles of formation the information and evaluation models of system and analysis of methods for finding useful information in them.

## 2. Complex hierarchical network structures and systems

The structure of information model of the system to some extent should reflect the structure of system itself. Among the most investigated system structures are networks and hierarchies. In this case, during studying the network structures usually do not take into account the presence of a certain ordering or subordination of components, although it certainly exists in the vast majority of network systems [15, 16]. On the other hand, hierarchical structures do not take into account the relationships between components of the same level of hierarchy [17]. At the same time, even in strictly hierarchical systems, such as the militaries, such connections exist. That is, the structures of real artificial and natural systems do not fit into the concept of "pure" network or hierarchy. Hierarchically-network structures, namely structures, each component of a certain level of hierarchy of which can be represented as a subnet of the lower level of hierarchy or we can subordinate him the such subnet, more accurately and naturally reflect the peculiarities of interactions in real complex systems [18, 19]. Such structures are most common in human society (public administrations, military and security services, industrial companies and religious denominations, etc.) and it is usually in them that the question of making decisions about further actions regarding the system arises most acutely [20]. In other words, the greatest interest in terms of decision-making theory is caused by complex hierarchical network systems (CHNS), formed on the principle of direct subordination [21]. Each system of this type can be divided into two main components. The first component is a network system of the lowest level of hierarchy, which usually implements the purpose of creation and existence of CHNS, ie provides the movement of certain types of flows - transport, resource, financial, information and so on. Therefore, this layer will be called the base system (Fig. 1). The second component is a hierarchically-network management system (MS), which with the help of management, organizational, information and other kinds of flows should ensure the effective functioning of base system, in particular, rapid response to both negative and positive changes, analysis and forecasting of these changes , timely prevention of the development of threatening trends [22], etc. In general, the CHNS of direct subordination is a multiflows multilayer system, the two main types of flows of which are concentrated at the lower and higher levels of the system hierarchy, respectively. Obviously, the existence of CHNS's management system without the base system it manages makes no sense. On the other hand, operation of the base network system without proper management often becomes virtually impossible. This is probably the main reason why the structures generated by such systems are called interdependent networks [23].

Consider, as an example of complex hierarchically-network system, the railway transport system (RTS) of Ukraine [14]. The defining feature of its consistent division into subsystems of lower hierarchical levels is a clear territorial principle of building the management system of Ukrainian railway. It includes 6 regional railways, 27 railway directorates, 110 track distances and more than 1,200 sections, which are usually a sequence of stations and inter-station races with a track length of 20 to 30 km. This principle of structuring makes it possible to establish a clear link between the components of RTS's base system and management units that are responsible for the state and operation process of these components. Further structuring is carried out on the basis of functional purpose, namely, as subsystems that form railway sections are distinguished such objects as stations (nodes) and interstation races (edges). That is, the railway transport system as CHNS has 6 levels of hierarchy. The flows in the base system are passengers and freight, and their carriers are passenger and freight trains, respectively.

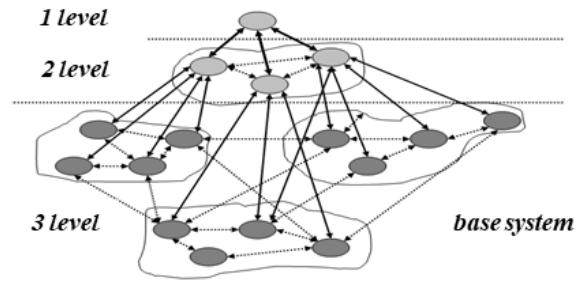

**Figure 1.** An example of a system with a three-level hierarchically-network structure of direct subordination

This method of structuring is used in most man-made industrial, financial, socio-economic, military, religious and other large complex systems. For example, the road network of a country, region, and locality. The nodes in such network are settlements, the edges are the roads connecting them, the flows are passengers and freight transported by vehicles.

Research and effective organization of real CHNS requires a holistic and complete picture of it. This picture is formed on the ground of all information about the history, current state and forecast of the behavior of such system. Based on this, we can form an information model of CHNS [18] ie identical to structure of system the dynamic (in sense of constant expansion and replenishment) data structure, each component of which contains information about the state and operation process of relevant system's component at present moment, past and future, beginning from the lowest level of hierarchy and ending with the system as a whole.

## 3. Usefulness of information

Before determining the principles of formation of information models of real CHNS, it is necessary to understand what data should be included in such models. Usually this understanding is reflected in the concept of "usefulness" of information for corresponding system. The usefulness of information is a rather subjective and relative concept. Ways to increase capacity and reduce the cost of batteries for smartphones, laptops or electric cars; new species of deep-sea fish; methods of reducing harmful emissions into the atmosphere; the history of Dogon tribe, the discovery of new mineral deposits - for each of these data there are systems for which they will definitely be useful. At the same time, important information for one of these systems may be completely irrelevant to others. There are data that are important for almost all people, such as the emergence of new effective drugs that can overcome cancer or other serious incurable diseases, and there are data that are useful only for separate, sometimes small systems, such as groups of people with rare diseases. The systems that arise from the need to overcome these diseases are different in the number of their elements and relationships (patients, doctors, manufacturers of appropriate equipment and medicines). That is, the usefulness of information, in addition to the subjective, has a certain objective or quantitative dimension, which is determined in particular by the "size" of system. However, quantitative usefulness indicators are often inaccurate or controversial. Information that seems irrelevant today, such as the latest mathematical and physical theories or the discovery of new chemical compounds, can become very important if not in the near, then in the distant future. Such were once considered the discovery of transuranic elements or X-rays, the development of graph theory or the study of superconductivity, and so on. These considerations imply the need to develop both global and local (for separete systems) usefullness criteria, also taking into account the potential importance of information that may be considered unnecessary at present. It is difficult to develop a universal criterion of usefulness, because it must reflect the importance of specific information for all mankind or at least for most of it. It is much easier to formulate usefullness criteria for separate systems. Without such criteria, it is difficult to develop reliable algorithms for finding useful information. This means that large amounts of necessary data can be lost. The usefulness of information for each system is manifested in its reliability, timeliness, clarity, aiding to achieve the goal of its formation and existence, the breadth of application in the present and future. By useful information we mean data that helps to better understand the processes that take place in the system, and can be used in decision-making that help optimize its state and operation or avoid threats that could destabilize or shut down the system. The fact that useful information was analyzed and on the basis of this analysis the right decision was made, is expressed in the increase of quality evaluations both separate elements of the system and its components at higher levels of hierarchy. In general, each system should form its own indicators and criteria for the usefulness of data that have to be included in information model of this system.

## 4. Information model of complex hierarchical network system

The main purpose of creating an information model is to direct on the appropriate levels and control elements of CHNS to simplify the process of data analysis and decision-making on further action in relation to their subordinate components of the system. Hence the natural requirements for the content, presentation, quality and volume of data, which fills the information model: objectivity, relevance, clarity, reliability, minimum sufficiency, completeness [5], etc. The main task that arises during the formation of information model of the system is the elimination of duplicate, unimportant and unreliable data and the structuring and storage of only useful information about the system. By structuring in this case, we mean the ordering of information on the basis of its belonging to the description of a particular component of CHNS in order to simplify further search and analysis (the data themselves can be both structured and unstructured).

Typically, the larger the volume of flows pass through separate element (node or edge) of the CHNS, the higher its priority in the system [24]. Therefore, it is advisable to calculate the quantitative measure of element priority, using the values of its parameters of influence and betweenness in CHNS [25]. Quantitative values of these parameters are determined by the volumes of flows generated by node and distributed by system, generated at other nodes of system and received at this node or passing through this node in transit. In general, influences can be both positive and negative and have different consequences. Thus, a strong electromagnetic field affects the health of population living near the generator of such field [26, 27] and the biochemical composition of water in reservoirs near it [28]. Equally important is the impact on environment of sources of radioactive contamination or harmful industries [29, 30], etc.

Elements with large values of parameters of influence and betweenness can be considered the most priority in the system and the data on these elements should be first of all included in the information model and be subject to priority research. The flow characteristics of nodes and edges of the CHNS's hierarchical levels make it possible to reasonably determine and dynamically change their priority in the system with changing conditions of its operation. The priority of subnet elements of a certain level of hierarchy is their local characteristic and does not determine the priority of element of the higher level of hierarchy to which they are subordinated. That is, an element may be a high priority in subnet, but the subnet itself may have a low priority among other subnets of a given level of hierarchy. On the other hand, the priority of a higher-level element of hierarchy often determines the priority of subordinated its lower-level elements.

To determine the level of fullness of the information model with necessary data, it is advisable to enter certain objective quantitative indicators that form the structure of fullness of this model [18]. The main criterion for fillness the data on a particular element or subsystem of CHNS is the ability to make on their basis a timely correct decision on further action regarding this element or subsystem.

Even following the formation principles described above, information models can contain incomplete data of extremely large volumes, which are unrealistic to process "manually" at reasonable intervals. One of the ways to solve this problem is to form on the basis of information model of CHNS the models of evaluation [14, 31]. Since objective and mathematically sound evaluation is possible only with the necessary information about the state, operation process and interaction of system elements of all levels of hierarchy, then collection, processing, storage and accounting of such information is one of the main goals of creating and filling the information model of researched system.

## 5. Evaluation models of complex hierarchical network system

Usually the cause of failures or inefficient functioning of the system is untimely or erroneous evaluation its current state or incorrect forecast of further development. Evaluation theory methods allow us to determine the prerequisites that can lead to such consequences in the work of CHNS, and to identify those elements that require urgent improvement of state, optimization of operation process or effective interaction with other components of the system [32, 33]. Each CHNS information model can be matched with a regular and / or interactive system evaluation model. The model of regular evaluation is based on information obtained during periodic planned surveys of CHNS or collected during a certain period of its operation, and provides (fig. 2) [32]:
• local evaluation of the state, quality of functioning and interaction of the elements of CHNS's base system;
• aggregated evaluation of system elements at all hierarchycal management levels, in which a generalized conclusion about the state and operation quality of the CHNS's subnet subordinated to a certain element is crucial for the evaluation of this system element;
• prognostic evaluation of the state and operation quality of elements of all levels of the CHNS hierarchy.

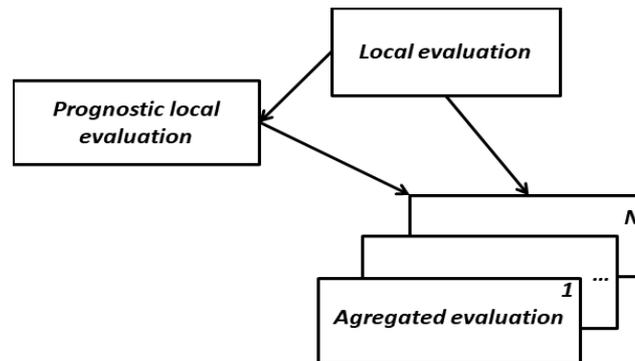

**Figure 2.** Scheme of system evaluation

The purpose of regular evaluation is a deep and thorough analysis of the state, process of functioning and interaction of all elements of the system. In this case, each of the characteristics of element, which are contained in the information model of CHNS, correspond a set of evaluations of its behavior on a certain set of criteria and parameters. The structure of aggregated evaluations is identical to the structure of assessed CHNS, and the choice of methods for constructing generalized conclusions (the weakest element, weighted linear or nonlinear aggregation, hybrid aggregation procedures) must take into account the specifics of a particular system [34]. Obviously, the period of forecasting cannot be shorter than necessary to overcome the identified threats.

The fillness and priorities structures in evaluation models determine the level of coverage of the CHNS elements of all levels of hierarchy, depending on the data available in the information model, and the priority of analysis of these evaluations, respectively.

The model of interactive evaluation of CHNS is based on the results of continuous monitoring of the system operation process, which are contained in the information model, and provides (fig. 2) [33]:
• local evaluation of the interaction of flows and structural elements (nodes and edges) in the CHNS's base system;
• aggregated evaluation of the interaction of system elements at all levels of hierarchy in which a generalized conclusion about the quality of interaction of elements subordinated to a particular element of the CHNS subnet is crucial for the assessment of this element;
• prognostic evaluation of the quality of interaction of elements of all levels of the CHNS hierarchy.

Interactive evaluation is performed continuously in real time and consists in constant monitoring of the interaction of network and interlevel flows with CHNS elements. During the continuous monitoring of large complex systems, significant amounts of information constantly come from many sources, which require the development of special methods for its operational processing on modern computers [35, 36]. The conclusions obtained from the interactive evaluation are indirect, but no less important for monitoring the state and operation quality of system and its separate components.

The model of interactive evaluation is formalized in the form of a dynamic data structure, which contains a set of characteristics of continuous interaction of flows moving CHNS, with elements of its structure and estimates of this interaction. The generalized conclusions received as a result of interactive evaluation for the period between regular researches of system, it is expedient to include in model of regular evaluation of CHNS. Such conclusions can adjust the sequence of the next planned study of system, namely, in each set of equally important elements, the regular investigations begins with the components that received the worst aggregated over time interactive evaluations.

The results of regular evaluation can also change the sequence of research of elements of real complex hierarchical network systems. If in the case of information models this sequence is determined by the structure of priority, then in evaluation models it naturally begins with the elements that received the worst evaluations, and therefore create the greatest threats to the work of at least related components of the system. In general, unlike information models, evaluation models contain only structured data. The main advantage of evaluation models is much smaller amounts of data, which are much easier to analyze and allow us to quickly locate the most threatening elements of the system. The use of such models significantly reduces the amount of information that needs to be analyzed in the first place, and thus is an effective means of overcoming the problem of quantitative complexity of systems research [37]. The identity of structures of the CHNS, its information model and evaluation models allows us to move from unsatisfactory evaluation of element to the data describing its state and operation process, analysis of this information to investigate the causes of shortcomings and make decisions direct on elimination such shortcomings in the system element.

Often the reasons for failures of system functioning or its separate components lie outside it: climate change and negative phenomena in wildlife are often the result of influence of industrial society, social unrest is usually caused by economic crises, production or sales planning must take into account existing of competitors etc. Taking into account and analysis of intersystem interactions and influences often allows us to identify these causes and prevent their consequences, which requires the development of tools for joint analysis of data contained in information models of interacting systems.

The application of evaluation theory methods is not limited to the search for poorly functioning components of the system. These methods are an effective tool for determining the exemplary elements that can serve as a practically achievable criterion of quality, finding the optimal modes of system operation [38] and choosing the optimal system from given class of equivalent systems [39]. Evaluation theory can be successfully used to identify objects or processes whose behavior goes beyond known standards or conceptions, or in other words, to novelty detection.

## 6. Novelty detection in information model of complex hierarchical network system

In the information model of the system during its replenishment may include various "anomalies", which are divided into two main types - the so-called outliers and novelty [40-42]. The search for outliers is to identify erroneous (measurement errors, errors in texts or numerical tables), as well as inaccurate, uncredible or duplicate data and their subsequent correction or deletion. Currently, to solve this problem, outliers are divided into separate types and to detect them, specialized software applications are being developed: programs for spell checking (Microsoft Word, Language Tool, AfterScan Express), programs for detecting duplication or plagiarism (Advego Plagiatus , ETXT, Easy Chair), programs for exposing fakes (RevEye, Tineye, Lauffeuer-Potential), etc. Removing such outliers from the IM of system can significantly lower the level of its "littering", reducing the time to find the necessary information, decreasing its volume and speeding up the process of analysis and decision-making.

The novelty detection is the purposeful identification of data that are not typical in one features or another for the relevant real system, incompatible with existing ideas and theories or those that do not meet new requirements and standards [43, 44]. Thus, some diseases have been known to doctors for decades: AIDS was detected in the 1920s in Africa, Zik's virus - in the 1950s in South America, SARS Cov virus - in 2002 in China, and MERS - in 2010 in the Middle East. That is, this information was known, but "neglected" by the medical community or relevant government agencies, despite the high mortality of these diseases. In general, the main feature of hidden novelty is the presence of relevant information in data warehouses, but not taking it into account when analyzing the system functioning. Usually, data that are a hidden novelty are not detected during regular research of the system or continuous monitoring of its operation process, because they are not the purpose of such research. Thus, during inspections of the railway track, it is possible to identify sections whose condition is close to unsatisfactory. However, in order to determine how often and quickly the state of such sections deteriorates during a long period of operation, it is necessary to analyze the data about these sections, which are contained in the IM of railway transport system of Ukraine. For example, in 2009, after a heavy downpour as a result of washed away the ground of railway track on the Lviv-Klepariv section, 9 freight train car derailed. As a result of analysis it turned out that similar washings in this place happened repeatedly. This information was contained in the railway database, but due to the lack of accidents it was "ignored". Only the emergency situation forced to appropriate strengthen the threatening section of railway.

Erroneous and new information is often formally similar. Thus, during evaluation of the quality of prosthetics of disabled persons [13], experimental data of changes in the angles in main joints of musculoskeletal system, which contained point "peaks" with an amplitude of $30^O$–$40^O$, were repeatedly encountered. It is obvious that the human body is unable to perform such movements, and therefore these peaks were the result of malfunctions of measuring equipment. At the same time, the results of defectoscope evaluation of the state of railway tracks [32] for the presence of cracks often contained data with such "peaks", which was evidence of a crack, ie important information that required prompt response (replacement of cracked rail). Of course, such peak in data can occur due to a malfunction of the flaw detector, but its presence under any circumstances requires an urgent response. In the 1960s, radio signals were received from space, which contained sharp periodic "bursts". At first they were perceived as failures of radio telescope, then - as condensed messages of extraterrestrial civilizations. Finally, after appropriate analysis, a new class of stars was discovered - pulsars or neutron stars. That is, one search criterion (the presence of "peaks" in the numerical data) can give both correct and erroneus, and controversial result, which requires more careful study.

In each information model it is necessary to allocate two main stages of data processing and analysis. The first of them consists in continuous monitoring and pre-processing of data entered into the IM of the system. At this stage, there is an analysis and elimination of definitely erroneus, unusable or duplicate information. Depending on the applied criteria and amount of data that can come continuously from many sources, one or another part of data with "outliers" enters the data warehouses, and data with present novelty can be rejected as erroneous. Information that contains an obvious threat to the system is subject to unconditional response. If such data enters the information model, then it is need to make operational decisions to counter the arised threat.

The data stored in the information model after pre-processing is the basis for finding the hidden novelty. The effect, which is determined by the novelty, can be manifested singly or en masse, be one-time or repeated regularly. To identify such effects, it is necessary to search the entire information model, ie in the whole set of available data about the system, which can be extremely many, of different quality and distributed among different components of the IM. Thus, the best results of prosthetics, which are taken as a practically achievable quality criterion [39], are searched in the whole database for a disabled persons with amputated lower limb, the track sections that due to adverse geological or climatic factors regularly lead to railway accidents - in the whole railway database [32], the highway sections on which road accidents occurred most often - in the whole road database, etc. The search in the whole information model is carried out in cases when operation process of the system changes or does not correspond to the expected due to reasons that are difficult to establish, for example, areas where the number of certain diseases significantly exceeds or life expectancy is much less than the known statistical norm.

Often the search for novelty should reveal the readiness or unreadiness of system or its separate components for new operating conditions. For example, the databases of railway transport system of Ukraine contain information on the current state and history of the results of evaluations of all track sections. These data allow us to determine the extent to which this state meet the conditions necessary for the introduction of high-speed trains on predetermined routes, by evaluating the behavior of current state of elementary track sections lying on the route, to the conditions necessary for the introduction of high-speed trains (seamless track, strengthened upper structure of track, increased vertical displacement of track threads on curved sections, etc.).

Another feature of the search for hidden novelty is the determination of its mass character, recurrence and distribution in space and time. An interesting example of such problem is the detection of actual number of diseased peaple with COVID-19. The search criterion in this case is a positive result of PCR- or ELIA-testing. Since about 80% of those infected carry the disease without any symptoms and often do not seek medical attention and not all citizens are tested because of its cost, the real picture of spread of coronavirus has not yet been established and is unlikely to be established in the future. That is, despite the presence of clear search criteria, it is impossible to form an objective conclusion about the real state of the system. A similar situation has developed with infections of hepatitis B and C, AIDS, etc., the course of which may be invisible for at least several years. At the same time, an infected person poses a real threat to others.

In order to detect the regularity and mass character of certain events in the system, it is often necessary to collect information for a sufficiently long period of time for all its elements. Thus, the side effects and contraindications are detected for some drugs. They are usually determined for a control group of patients, although a complete picture requires an analysis of action of this drug for all those who took it, and even subsequent generations, as happened with patients who took the drug thalidomide, or those who had Zika's disease. Many other harmful influences are invisible and manifest their effects over long periods of time (insignificant, at first glance, but constant influences of electromagnetic fields, small doses of radiation or chemicals, false information, propaganda, etc.).

As follows from the above examples, the "novelty" of data stored in the IM can be different, sometimes quite specific character. In order to solve this problem, methods of data mining, machine learning, neural networks, etc. are currently being developed. Evaluation theory methods can also be successfully used in problems of finding hidden novelty by means of creation the novelty evaluation models. The development of such models is greatly simplified by the numerical format of data stored on electronic media. Thus, the processing of audio file for voice identification or video file for recognizing a particular person's face can be quite effectively performed using local evaluation algorithms (evaluating the deviation of numerical image of an audio or video sequence from a given sample [32]). To generalize the obtained search results, as well as the formation and prediction of conclusions about their mass character and repeatability, it is advisable to use methods of nonlinear and hybrid aggregation, as well as known forecasting algorithms [34]. The identified novelty, especially if it has regular and mass character, should be included, as a subject of research, in the models of regular and interactive evaluation of the system. In combination, the interconnected models of regular, interactive and novelty evaluation form a model of complex evaluation (MCE), the purpose of which is the completeness and versatility of system investigation (Fig. 3).

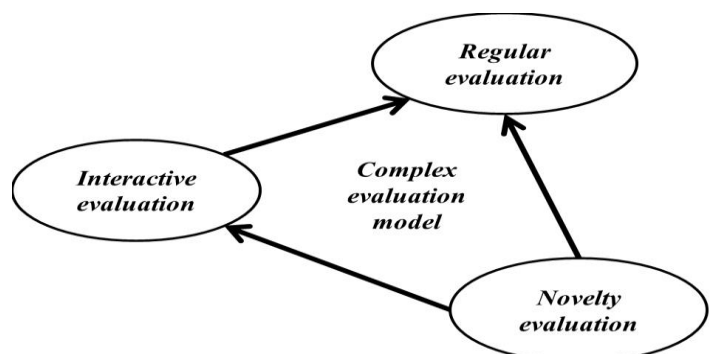

**Figure 3**. Scheme of complex evaluation model

Creating the model of complex evaluation sets quite strict requirements for information model of the system: structuredness, careful data pre-processing, regular cleaning of information "garbage" and so on. Obviously, the forming of system IM that meets these requirements requires considerable effort, but the further benefits of a much simpler, faster, and more effective search for the nesessary information definitely justify that effort. Searching for the necessary data in the IM and MCE of the investigated system, and prompt processing of this data in order to make appropriate decisions requires the use of the latest computer technologies for high-speed processing of huge amounts of information.

# 7. Optimization of data processing in information models of complex systems

As mentioned above, a complex methodology [14] has been proposed to analyze the state and operation process of CHNS, combining methods of local, prognostic, aggregated and interactive evaluation of system components. This technique is based on the use of large amounts of input data, which in some way characterize the elements of system and/or the system as a whole, and involves a number of parameters and evaluation criteria, as well as different operation modes. Usually the input data comes continuously, has various types (numerical, text, video, audio, image, etc.), can be poorly structured or unstructured, very often inaccurate, distorted or damaged. Therefore, they need to be pre-processed quickly before further use in evaluation procedures. In addition, investigation of most CHNS must be conducted in real time [12]. Therefore, to solve these problems it is necessary to develop effective parallel methods and computation algorithms, focused on the implementation on modern and perspective computing tools – computers with multi-core processors, clusters [45], hybrid architectures (clusters with graphic or quasisystolic processors for performing the special computations), high-performance computing environments [6], etc.

A quasisystolic digital filtering method has been proposed for real-time pre-processing of input data [35, 36], which was developed for the case of using the adaptive smoothing procedure [46]. Based on this method, the optimal in terms of speed and memory usage parallel-pipeline algorithms (PPA) for solving filtering problems of different dimensions were developed. Optimality is proved in the class of algorithms equivalent for the information graph. Software simulation of digital filtering PPA operation on a computer with a multi-core processor has been performed. The simulation results confirmed the correct operation of these algorithms. The constructed PPA filtrations are focused on the implementation on quasisystolic structures and computing means with structural and procedural organization of computations. Highly parallel digital filtering algorithms have been developed [47], which use a synchronous calculation scheme and a pyramid method [48] for parallelization of loops. These algorithms are focused on implementation on widely available means for users – computers with a multi-core processors and clusters.

A general approach [45] to the parallelization of methods of complex evaluation of CHNS is proposed, which provides for the simultaneous execution of blocks of calculations from given sets. That is, for each set of blocks is a large-blocks parallelization. Depending on the investigated object of the system and the ultimate goal of evaluation, all or only some of the above sets can be considered. Note, that each block implements only one of the types of evaluation (local, prognostic, aggregated, interactive) or explores only one of the categories (state, functioning, interaction) for a given object (element, subsystem, system). To further reduce the computation time, parallelization are implemented within each of the blocks.

A parallel-sequential approach [49] has been developed to optimize calculations in the case of local evaluation of the CHNS element characteristic according to the given parameters. Note, that when calculating the value of local evaluation parameter, a mode close to the full binary tree was used. For separate fragments of calculations, efficient algorithmic constructions have been developed, focused on implementation on universal parallel computing means with shared memory. In [31], the efficient algorithmic constructions for parallelization of computations during obtaining generalized conclusions (aggregated evaluation) for different components of CHNS on universal parallel computing means with shared and distributed memory are proposed. The procedure of interactive evaluation of CHNS components is formalized and for its realization the corresponding algorithmic constructions are proposed, which reveal the perspective of parallelization and take into account the limited possibilities of computing resources [50]. Estimates of complexity and acceleration of parallel calculations are given, which confirm the high efficiency of using these constructions. Effective algorithmic constructions have been developed to predict both the evaluations themselves and the behavior of characteristics of system components on parallel computing means with shared and distributed memory. Software implementation of separate algorithmic constructions for parallel organization of procedures of evaluation of CHNS components on the computer with multi-core processor is carried out. The results of numerical experiments confirmed the effectiveness of this implementation.

Along with obtaining adequate quantitative evaluations for CHNS components, the problem of visualization of evaluation results is no less important. To solve it, original approaches have been developed [32, 33], which allow to see not only quantitative but also qualitative picture of system behavior. In this case, the visualization itself can be performed in real time simultaneously on a number of widescreen monitors connected to a computer network or on separate its node. Each of these monitors can reflect a number of windows with evaluation results. Therefore, at the same time it is possible to obtain and analyze a significant amount of qualitative information about the state and operation process of CHNS components. In summary, we can state that the developed parallel methods and algorithms for processing large amounts of information during investigation of complex systems allow us to some extent solve the problem of Big Data in real CHNS, including transport networks, water and energy supply systems, etc.

## 8. Conclusions

The continuous and rapid growth of data amounts generated by human society in various areas of its activities places new demands on technologies for storing and processing this information. The problem of information overload is becoming more acute, which creates the phenomenon of "paralysis of analysis" and nullifies further attempts to comprehensive informatization of society. The article proposes approaches to at least a partial solution of this problem, which are to create information models of the most important for human life complex systems and build on their basis models of complex evaluation of such systems. These models and means of operational data processing allow us to analyse in real-time the current behavior of the system and prospects for its further development and are a powerful mathematical tool for creating expert systems and decision support systems in various subject areas.